\documentclass[12pt,a4paper,final]{iopart}

\usepackage{iopams}  
\usepackage{graphicx}
\usepackage{etoolbox}
\apptocmd{\thebibliography}{\raggedright}{}{}
\usepackage[breaklinks=true,colorlinks=true,linkcolor=blue,urlcolor=blue,citecolor=blue]{hyperref}
\expandafter\let\csname equation*\endcsname\relax
\expandafter\let\csname endequation*\endcsname\relax
\usepackage{amsmath}
\usepackage[caption=false]{subfig}
\usepackage{enumerate}

\begin{document}
\bibliographystyle{vancouver}

\title[High-Pressure CO$_2$ Dissociation]{High-Pressure CO$_2$ Dissociation with Nanosecond Pulsed Discharges}

\author{Erwan Pannier$^{\footnotemark,1,2}$}
\address{$^1$Stanford Plasma Physics Laboratory, 452 Escondido Mall, Bldg. 520-118, Stanford, California 94305, USA}
\address{$^2$Laboratoire EM2C, CNRS, CentraleSup\'{e}lec, Universit\'{e} Paris-Saclay, Grande Voie des Vignes, 92295 Chatenay-Malabry cedex, France}
\ead{erwan.pannier@centralesupelec.fr}

\author{Taemin Yong$\ddagger$}
\address{Stanford Plasma Physics Laboratory, 452 Escondido Mall, Bldg. 520-118, Stanford, California 94305, USA}
\ead{taemin@stanford.edu}

\footnotetext{E.P and T.Y. contributed equally to this work..}

\author{Christophe Laux}
\address{Laboratoire EM2C, CNRS, CentraleSup\'{e}lec, Universit\'{e} Paris-Saclay, Grande Voie des Vignes, 92295 Chatenay-Malabry cedex, France}
\ead{christophe.laux@centralesupelec.fr}

\author[cor1]{Mark A. Cappelli}
\address{Stanford Plasma Physics Laboratory, 452 Escondido Mall, Bldg. 520-118, Stanford, California 94305, USA}
\ead{cap@stanford.edu}

\begin{abstract}

The efficiency of the conversion of  CO$_2$ into CO with nanosecond repetitively pulsed discharges (NRP) is investigated in a high pressure batch reactor. Stable discharges are obtained at up to 12~bar. By-products of  CO$_2$ splitting are measured with gas chromatography. The energy efficiency is determined for a range of processing times, pulse energy, and fill pressures.
The energy efficiency is found to be approximately 20\% and is only weakly sensitive to the plasma operating parameters, \textit{i.e}, the extent of CO$_2$ conversion is almost linearly-dependent on the specific energy input. A conversion rate of up to 14\% is achieved with an energy efficiency of 23\%. For long processing times, a drop in efficiency is observed, due to the increasing significance of recombination reactions, as described by a macroscopic kinetic mechanism. Reaction pathways that are  believed to play  an important role in nanosecond pulsed discharges are discussed. It appears that vibrational excitation does not play a significant role in CO$_2$ conversion in these types of short-pulse discharge. Results also draw attention to the relative importance of two particular electronic excitation reactions. 

\end{abstract}


\vspace{2pc}


\section{Introduction}

The splitting of CO$_2$ and its conversion into a combustible chemical feedstock is one of several possible methods that can be used to store electrical energy that is generated from renewable resources during off-peak demand periods. The CO that is produced can be mixed with hydrogen to create synthesis gas which is a compact, high energy-to-mass-ratio form of chemical energy storage~\cite{Lebouvier2013,Goeppert2014,Petitpas2007}. The main challenge of this reforming process is the requirement to dissociate  CO$_2$ with a high energy efficiency,  ($\eta$) \textit{i.e.}, the ratio of the dissociation enthalpy to the electrical energy invested, and a high conversion degree, {i.e.}, the ratio of the number of CO molecules formed to that in the  original feedstock. For that, non-equilibrium plasma discharges are particularly suited as they are not bound by equilibrium chemistry. It has been suggested that low-pressure microwave plasma discharges can have energy efficiencies that may be as high as 61\%~\cite{Rusanov1981} for conversion degrees of about 10\%. Recent experimental studies~\cite{VanRooij2015,Goede2014} and modeling~\cite{Kozak2015} of microwave discharges report on energy efficiencies of up to 50\% and confirm the importance of vibrational excitation in the dissociation kinetics. Other low pressure plasmas such as radio-frequency (RF)~\cite{Spencer2011d} and DC glow discharges~\cite{Wu1996, Wang1999}, have been used to study CO$_2$ splitting with varying degrees of energy efficiencies reported, but generally less than 15\%.   

While low pressure plasma discharges offer more favourable conditions to exploit nonequilibrium electron energy conditions, high pressure discharges (such as those at atmospheric pressure or higher) may be more suitable for reforming applications because of their higher throughput and lower capital cost. However, the energy conversion efficiency for microwave discharges appears to decrease with increasing pressure~\cite{Rusanov1981}. Studies carried out at atmospheric pressure show an efficiency limited to 20\%~\cite{Spencer2012a} for a conversion degree of 10\%. This decrease in efficiency is explained by the increase in losses due to vibrational-translational (VT) relaxation~\cite{Kozak2015} at higher pressure. Recent results~\cite{Mitsingas2016} suggest the energy efficiency can be greatly improved (\textgreater90\%) but at the cost of a low conversion degree (2\%), which impairs the higher throughput obtained with increased pressure. Other atmospheric pressure techniques have been investigated, such as dielectric barrier discharge (DBD) splitting~\cite{Paulussen2010,Brehmer2014,Aerts2015}, which have reported conversion degrees of up to 35\% but with energy efficiencies limited to 5-8\%. Using catalysts and packed-bed reactors~\cite{VanLaer2015, Mei2015d,Zeng2015c}, the energy efficiency of DBD splitting improves slightly, to 10\%, while achieving a conversion degree of 20\%. Other discharge splitting strategies have been reported, including the use of CO$_2$-N$_2$ or CO$_2$-Ar gas mixtures~\cite{Snoeckx2016, Ramakers2015d} but the additives were not found to have a significant effect on the energy efficiency. In contrast, however, atmospheric pressure gliding arcs~\cite{Indarto2007,Nunnally2011, Tu2014} were reported to reach efficiencies comparable to those measured using sub-atmospheric microwave discharges ($\eta$=43\%) with a conversion degree of 10\%~\cite{Nunnally2011}.

Another promising candidate for the splitting of CO$_2$ at high pressures is nanosecond repetitively pulsed (NRP) discharges. Their short pulse width (10~ns) and low duty cycle (1/1000) makes them potentially energy efficient as most of the discharge energy resides in the electron fluid, which can participate in chemistry before thermalization with the background gas. These discharges are also scaleable in pressure and can be operated at conditions higher than atmospheric pressure for even greater reactant throughput. Studies of the dissociation of CO$_2$ in CO$_2$/CH$_4$ mixtures using NRP discharges has been  previously reported~\cite{Scapinello2016}. However, only one prior study exists of the splitting of pure CO$_2$ at atmospheric pressure or higher. This study was carried out by Bak \textit{et al.}~\cite{Bak2015a} in a flowing reactor for pressures ranging from 1 to 5~bar.  In that study, conversion energy efficiencies were reported to be as high as 11.5\%. 

In the present study we extend the NRP discharge splitting of CO$_2$ to even higher pressure. We show that creating and maintaining a nanosecond pulsed discharge under very high pressure can be a challenge. We describe the use of a chemical process to manufacture electrodes suitable for breakdown at high pressure. An analysis of the conversion chemistry is extended to 12~bar, with the quantification of the produced by-products using gas chromatography. Finally, we use these measurements to determine the energy and molecular conversion efficiency of the process and compare our results to those of prior studies.

\section{Experimental setup}

\begin{figure}[hp]
\centering
\includegraphics[width=4.5in]{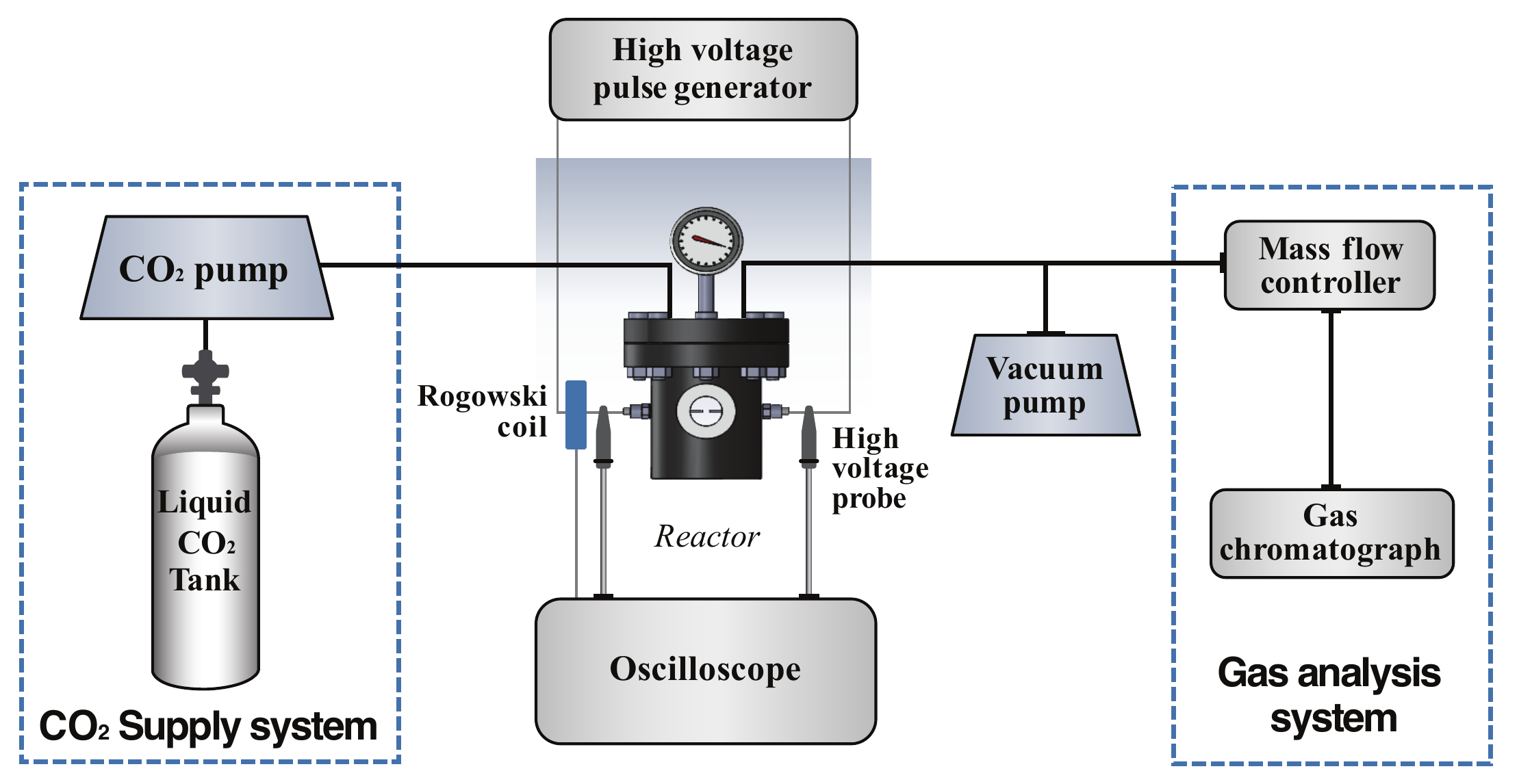}
\captionsetup{justification=centering,margin=2cm}
\caption{Process flow diagram for the high pressure pulsed plasma discharge splitting studies.}
\label{figure1}
\end{figure}

The high pressure facility consists of a reactor chamber, a high voltage pulse generator, a high pressure CO$_2$ supply system, a temperature control system, a gas chromatograph (GC), and a vacuum pump (Fig.~\ref{figure1}). The high voltage pulse generator (FID Technology Model F1112) delivers two signals of opposite polarities at up to $\pm$15~kV, with a duration of approximately 10~ns (FWHM) and a 5~ns rise time. The pulse repetition frequency is variable, but is held constant at 30~kHz for all the experiments reported here. 

The reactor chamber is designed to sustain high pressures of up to 12.5~MPa. Sapphire windows are built into the chamber for optical access. Two feedthroughs provide electrical access for generating the discharge. The inner volume of the chamber, as designed, was 400~cm$^3$, but then reduced to 200~cm$^3$ by filling inert material to raise the conversion degree of CO above the detection threshold of the GC, as needed for the case of higher pressures. 

\subsection{Electrode fabrication process}
 
The electrode configuration consists of a pin-to-plane geometry. The pin electrode has a radius of curvature of about 10~$\mu$m. It is fabricated from tungsten wire using an electrochemical etching process that we describe in detail in Appendix A. The gap distance is about 150~$\mu$m. The electrode length is known to within an accuracy of 30~$\mu$m resulting in an equivalent uncertainty in the gap distance. 

\subsection{Product gas characterization}

The product gases are analysed by a Varian (Model 3400) GC, equipped with two columns (Molecular Sieve 5Å and Porapak Q), and two non-destructive detectors in series (a thermal conductivity detector (TCD) and a pulsed discharge helium ionization detector (PDHID)).  The molecular sieve column can separate molecular oxygen (O$_2$) from molecular nitrogen (N$_2$), which allows us to ensure that there are no leaks in the system. Both the TCD and PDHID can detect O$_2$, CO and CO$_2$, but have a different sensitivity ranges. The TCD has a 1\% resolution while the PDHID can detect mole fractions down to 100~ppm. In practice, a mole fraction of 0.5\% is necessary in order to get reasonable quantitative data on CO. Appendix B details how we determine the conversion degree and number of CO molecules produced from the GC signals, taking into account the compressibility of CO$_2$ at high pressures. 

\subsection{Discharge energy measurements}
\label{section_energy}

Discharge energy measurements are facilitated by the use of two high voltage probes (Tektronix P6015A, 75~MHz range) and a Rogowski current probe (Pearson 2877, 200MHz, 2~ns rise time) connected to a Digital Oscilloscope (Tektronix TDS7104, 1GHz). Energy measurements require synchronizing the probes with 0.1~ns precision. This is done by applying a high voltage pulse with a voltage $V < V_b$, \textit{i.e.}, less than the breakdown voltage of the CO$_2$ within the chamber. Synchronization depends on the response time of the different probes and the length of the cables, which were kept constant for all of the experiments reported on here. 

\begin{figure}
\centering
\includegraphics[width=4in]{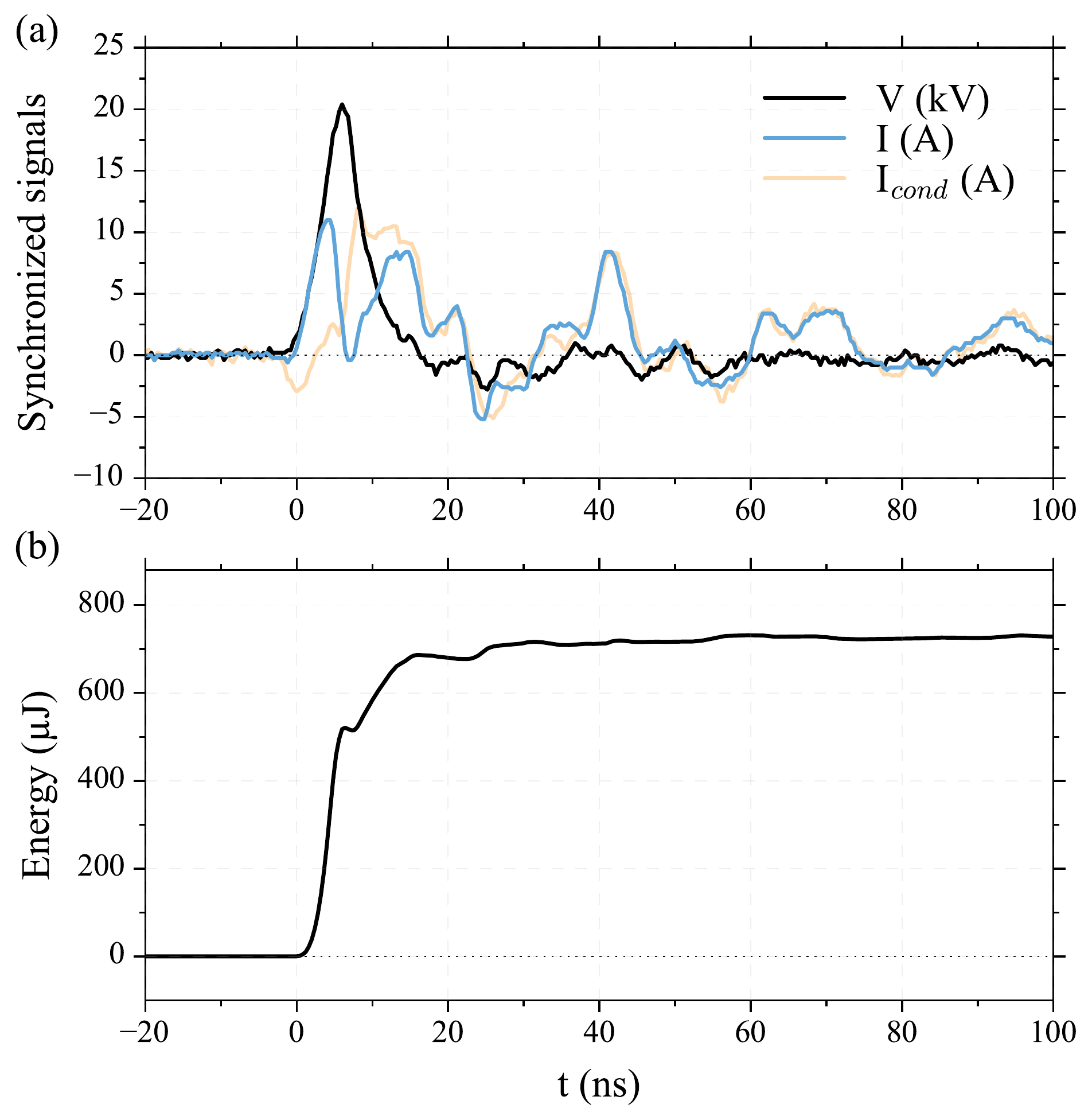}
\captionsetup{justification=centering,margin=2cm}
\caption{Typical signals in a NRP discharge (5~bar, 150~$\mu$m gap). (a) Synchronised voltage (kV), total current and conduction current (A). (b) Energy ($\mu$J)}
\label{figure2}
\end{figure}

Fig.~\ref{figure2} (a) shows the typical current and voltage traces seen during a NRP discharge event in CO$_2$ at 6~bar, when a 20~kV (peak) voltage was applied between the two electrodes. The peak voltage, V$_{max}$, is lower that the open circuit voltage, V$_0$ (not shown) as the electrical resistance of the gap drops as the gas breaks down, a process which begins (onset of conduction current, I$_{cond}$) prior to the application of the maximum voltage. The total energy deposited is measured to be 730 $\pm$ 50 $\mu$J (Fig.~\ref{figure2}, b). The conduction phase starts after about 5~ns, when the total discharge current peaks at 11~A, and the conduction current eventually rises to 12~A (the displacement current, not shown, is then negative). A large fraction (95\%) of the energy is deposited into the gas within the first 7~ns of the conduction phase. After 12~ns, the gap resistance is estimated to be $<$ 10~$\Omega$ and the voltage drop across the gap is small. The residual voltage and current peaks distributed regularly every 30~ns correspond to reflections of the initial pulse ringing in the 3~m transmission line. Although the total current is still fairly high, the voltage across the gap is  too low (because of the relatively low plasma resistivity) to deposit any additional energy, and the total cumulative energy deposited within the discharge levels off (lower plot).

\begin{figure} 
\centering
\includegraphics[width=4in]{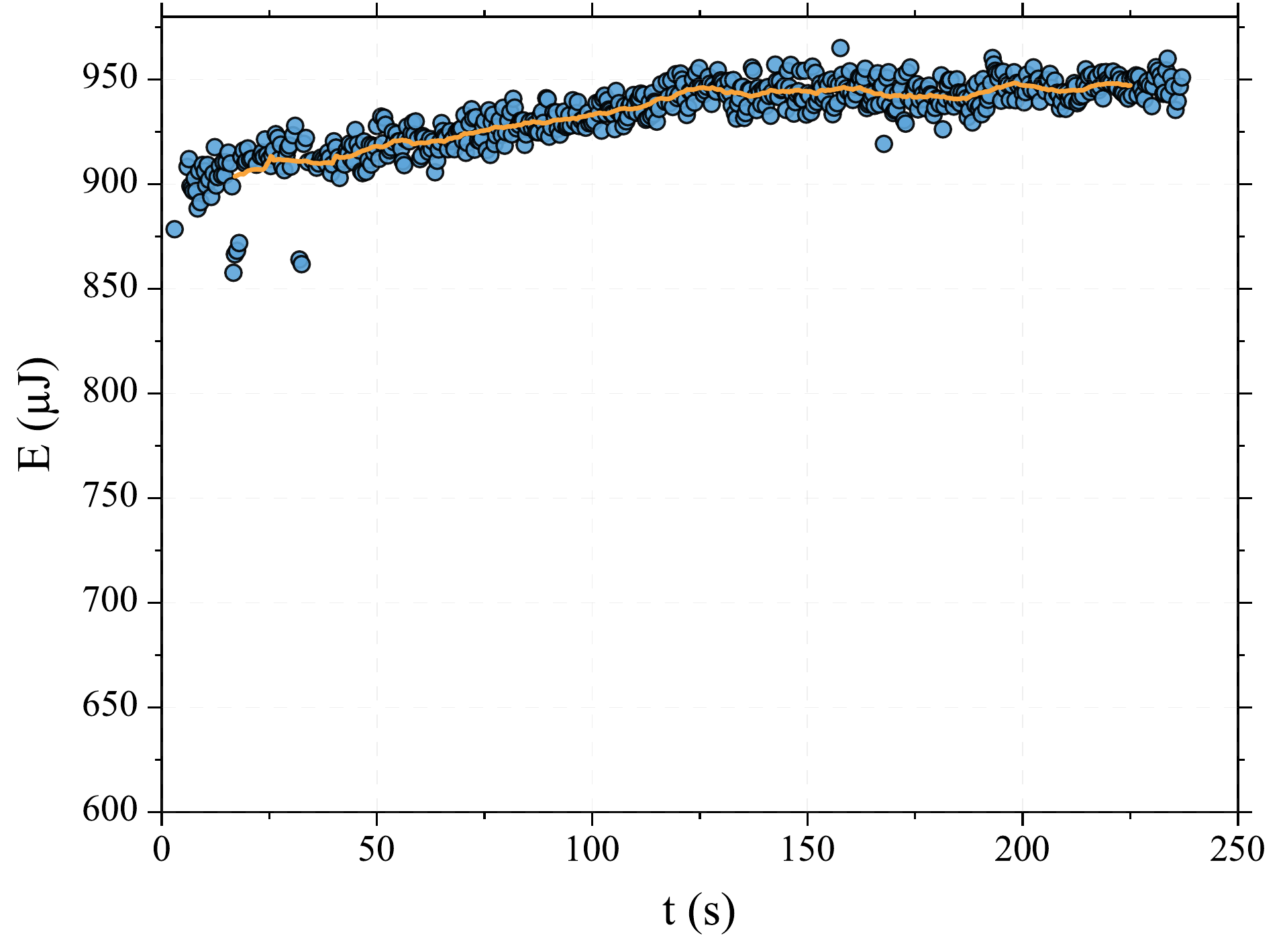}
\captionsetup{justification=centering,margin=2cm}
\caption{Energy fluctuation in the NRP discharge. p~=~5~bar, f~=~30~kHz, V$_0$~=~30~kV. Average energy per pulse: E~=~935~$\pm$~10 $\mu$J.}
\label{figure3}
\end{figure}

The total deposited energy is found to increase slightly with each new pulse over the course of an experiment of duration (processing time) of about 240~s. An example of this drift is shown in Fig.~\ref{figure3}, for the case of a 5~bar CO$_2$ plasma with pulses of a maximized, constant open-circuit peak voltage V$_0$ = 30~kV. Shot-to-shot fluctuations in the energy measurement are of the order of 10~$\mu$J, which, together with the uncertainty of the energy measurement of any single pulse (\textit{e.g.}, 50~$\mu$J in Fig.~\ref{figure2}), comprises the total uncertainty in the pulse energy. 

We see that the energy per pulse increases over time by an amount of almost 50~$\mu$J in a span of 240~s. We attribute this drift to the slow degradation of the electrodes. The gap distance and the radius of curvature of the pin electrode increase, resulting in a lower electric field and higher impedance. We measured an increase of the gap distance from 150~$\mu$m to about 250~$\mu$m in a time span of 240~s. When the applied voltage is close to the breakdown voltage condition, the electric field may decrease below the initial breakdown field, but the reduced electric field E/N is maintained as the high frequency induces residual heating. 

The pulse energy can be adjusted by varying the pulse voltage, as increasing the voltage increases the discharge current. Since the breakdown voltage increases with pressure, the minimum pulse energy needed for breakdown also increases. If the pressure is too high, this results in a fast ablation of the electrodes within a few cycles, hence the plasma cannot be sustained. With the electrodes designed for this study, stable discharges lasting sufficiently long to acquire good conversion data were obtained to pressures as high as 12~bar.

\section{Results and discussions}

\subsection{Energy efficiency}
The conversion energy efficiency for CO$_2$ dissociation is defined as~\cite{Fridman2008j}:
\begin{equation}
\eta=\frac{\Delta H}{E_{CO}}
\label{eq1}
\end{equation}
Here, $\Delta$H is the enthalpy of the CO$_2$ dissociation reaction:
\begin{equation}
CO_2 \rightarrow CO + \frac{1}{2}O_2, ~~~ \Delta H=2.9~eV
\label{eq2}
\end{equation}
and $E_{CO}=E_{tot}/N_{CO}$ is the average energy needed for each CO molecule that is produced, the total number of which is $N_{CO}$. The total energy deposited in any single experiment, $E_{tot}$ is given as
\begin{equation}
E_{tot}=E_{pulse}\cdot f\cdot\tau
\label{eq3}
\end{equation}
\textit{i.e.}, it is the product of the pulse frequency, f , the processing time, $\tau$, and the energy deposited per pulse, $E_{pulse}=\int_{pulse}{V I}dt$. In our experiments, the frequency is held constant at 30~kHz, the processing time varies from 30~s to 7~min, and $E_{pulse}$ varies from 300~$\mu$J to 1~mJ.

The total number of CO molecules produced, $N_{CO}$ , is derived from the gas chromatographic measurements. Appendix B describes the relationship between the number of molecules and the GC signals, taking into account CO$_2$ compressibility and the effect of the presence of other residual molecules not detected by the GC. The conversion efficiency, $\Phi$, is defined as the ratio of the number of CO molecules produced to the initial number of  CO$_2$ molecules in the reactor, $N^0$:
\begin{equation}
\Phi=\frac{N_{CO}}{N^0}.
\label{eq4}
\end{equation}
In the literature, the conversion degree is sometimes referred as the (molecular) conversion efficiency~\cite{Spencer2012a}. In Appendix B, it is shown that the conversion degree is expressed as $\Phi=\chi_{CO}/(\chi_{CO}+\chi_{CO_2})$ where the mole fractions of CO and CO$_2$ ($\chi_{CO}$ and $\chi_{CO_2}$, respectively) are a direct output of the GC measurements. The initial number of CO$_2$ molecules, $N^o$ , is calculated from the initial pressure and temperature in the chamber ($P_0$ and T), the chamber volume, $V$, and the CO$_2$ compressibility, $Z(P,T_0)$~\cite{Spencer2012a}. The number of CO molecules produced is then:

\begin{equation}
N_{CO}=\frac{\chi_{CO}}{\chi_{CO}+\chi_{CO_2}}\cdot\frac{P_0V}{Z(P,T_0)RT}
\label{eq5}
\end{equation}
The energy efficiency, as defined in Eqn. (\ref{eq1}), can  be calculated from Eqns. (\ref{eq2}) and (\ref{eq4}). For this purpose, it is useful to introduce the specific energy input (SEI); 
\begin{equation}
SEI=\frac{E_{tot}}{N^0}
\label{eq6}
\end{equation}
The SEI corresponds to the mean energy used per CO$_2$ molecule that was present initially in the batch reactor, or in the inlet flow for the case of a flowing system.  The SEI is a parameter that allows us to compare different plasma dissociation techniques. Eqn. (\ref{eq1}) can be rewritten in terms of the SEI:

\begin{equation}
\eta=\frac{\Delta H\cdot\Phi}{SEI}
\label{eq7}
\end{equation}

\subsection{Parametric study}

\begin{figure}[hp]
%
\centering
\includegraphics[width=4in]{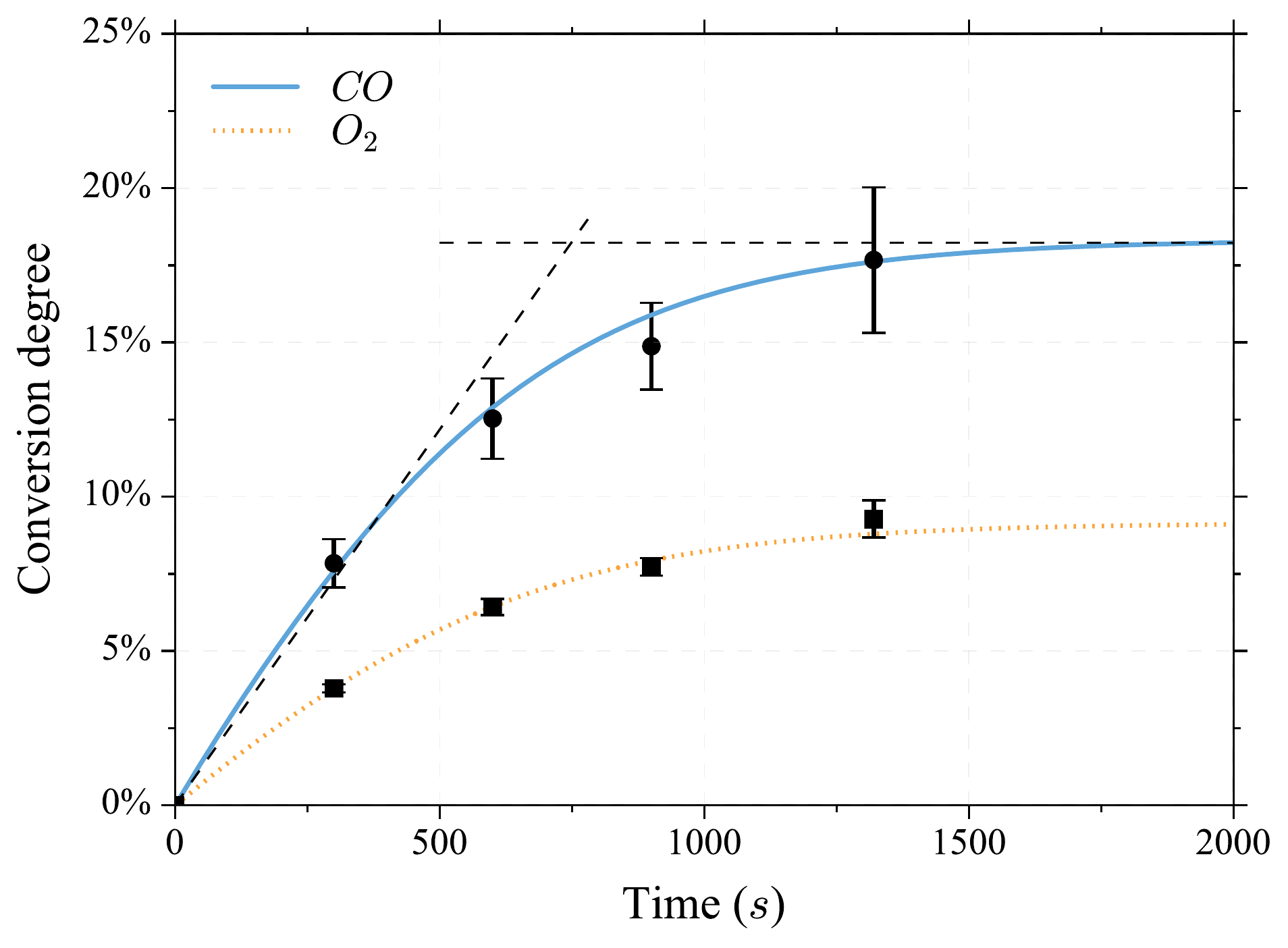}
\captionsetup{justification=centering,margin=2cm}
\caption{Evolution of conversion degree with processing time (p~=~5 bar, f~=~30~kHz, U~=~700~$\mu$J).}
\label{figure4}
\end{figure}

Experiments were performed to assess the dependency of the conversion degree on the processing time. The input energy was maintained constant at 700~$\mu$J per pulse and the pulse repetition rate was 30 kHz (21 W). The results are shown in Fig.~\ref{figure4} as solid circles.  The experiments are not carried out in any particular order (in terms of processing time) so as to prevent any systematic bias associated with potential electrode erosion.
During the first 400~s of the experiment (to the point where the conversion degree reaches about 10\%) the conversion degree increases almost linearly with time. Beyond this initial period we observe the onset of a saturation in the conversion. This saturation is attributed to the reactions that destroy the formed CO. The conversion appears to follow an exponential growth of the form $\Phi=\Phi_{max}\left(1-exp(-Kt)\right)$, with $\Phi_{max}\sim18\%$. This trend can be explained by a simple four-step kinetic mechanism:

\begin{align}
e+CO_2 &\xrightarrow{k_1} e+CO + O \label{eq8}\\
O+CO_2 &\xrightarrow{k_2} CO + O_2 \label{eq9}\\
O+CO &\xrightarrow{k_3} CO_2\label{eq10}\\
O_2+CO &\xrightarrow{k_4} CO_2 + O\label{eq11}
\end{align}

The first step in the mechanism is the direct electron impact dissociation reaction, assumed to be described by a bimolecular rate constant, k$_1$. The second step is an oxygen abstraction reaction that forms stable species CO and O$_2$, with a rate constant, k$_2$. The third step is the recombination of CO and O to regenerate CO$_2$, described by rate constant k$_3$.  Finally, the fourth step is the reverse reaction of step 2, which may become important as the CO and O$_2$ levels rise, and is responsible for the saturation seen in the experiments with a rate constant, $k_4$.

We can express the rate of change with time of the number of O, CO, and O$_2$ molecules formed as:

\begin{align}
\frac{dN_O}{dt}&=k_4N_{O_2}N_{CO}+k_1n_eN_{CO_2}-k_2N_ON_{CO_2}-k_3N_{CO}N_O\label{eq12} \\
\frac{dN_{CO}}{dt}&=-k_4N_{O_2}N_{CO}-[(k_3+k_2)N_O+k_1n_e]N_{CO}+(k_2N_O+k_1n_e)N^o\label{eq13}\\
\frac{dN_{O_2}}{dt}&=-k_4N_{O_2}N_{CO}+k_2N_ON_{CO_2}\label{eq14}
\end{align}





Here, $n_e$ is simplified to be the quasi-steady (reactor volume-averaged) electron density and is taken to be constant, independent of time. A carbon balance is used to relate $N_{CO_2}$ to $N_{CO}$ and the initial total number of molecules $N^o$, \textit{i.e.}, $N_{CO_2}=N^o-N_{CO}$. Together with Eqns.~(\ref{eq12}-\ref{eq14}), the carbon balance closes the equation set which can be solved, assuming values for the reaction rate constants, the electron density, and the initial number of CO$_2$ molecules in the reactor. 

The characteristics of this model agree qualitatively with the experimental measurements.  If the electron density and initial number of CO$_2$ molecules are known, then the model yields the rate constants, $k_j$. The solid (blue) curve in Fig.~4 shows the result of the model, assuming a quasi-steady reactor volume averaged electron number density of $n_e$ = $10^{16}$ $m^{-3}$, and an initial number of molecules $N^o = 4.88 \times 10^{22}$. The extracted values from this model gives $k_1 = 1.44\times 10^{-20} m^3 s^{-1}$, $k_2 = 2.23 \times 10^{-23} s^{-1}$, $k_3 =4.23\times 10^{-23} s^{-1}$, and $k_4 = 3.38\times 10^{-25} s^{-1}$.  Also shown in Fig.~4 is the measured $O_2$ conversion degree (solid square symbols), defined as the ratio of the number of $O_2$ molecules formed relative to the original number of $CO_2$ molecules in the reactor, $N^o$.  The dashed brown line is $N_{O_2}$/$N^o$ that is predicted by the model. We see that when rate constants are adjusted to get quantitative agreement with the evolution of $N_{CO_2}$, we also see good agreement with the measured $N_{O_2}$. In the above mechanism, we assume that no other products of CO$_2$ splitting are formed in significant numbers. This assumption implies that for long processing times, $N_{O_2} \approx \frac{1}{2} N_{CO}$. Previous research using  atmospheric DBD plasmas~\cite{Brehmer2014} have shown that additional oxygen compounds are a likely by-product of CO$_2$ dissociation, in particular, ozone (O$_3$), which is the product of the recombination of atomic and molecular oxygen. However, the residual heating induced by our discharge is expected to reach values comparable to NRP discharges of similar energies in air~\cite{Rusterholtz2013,Pai2010r}, \textit{i.e.}, at least 500~K, which is likely to prevent O$_3$ formation. The assumption of the negligible formation of other oxygen products is supported by our experimental measurements. The gas chromatographic measurements confirmed that $N_{CO}\sim2\,N_{O_2}$ in all cases studied, and an oxygen balance would yield ozone concentrations much smaller than the estimated uncertainty of our measurements. 

\begin{figure}
\centering
\includegraphics[width=4in]{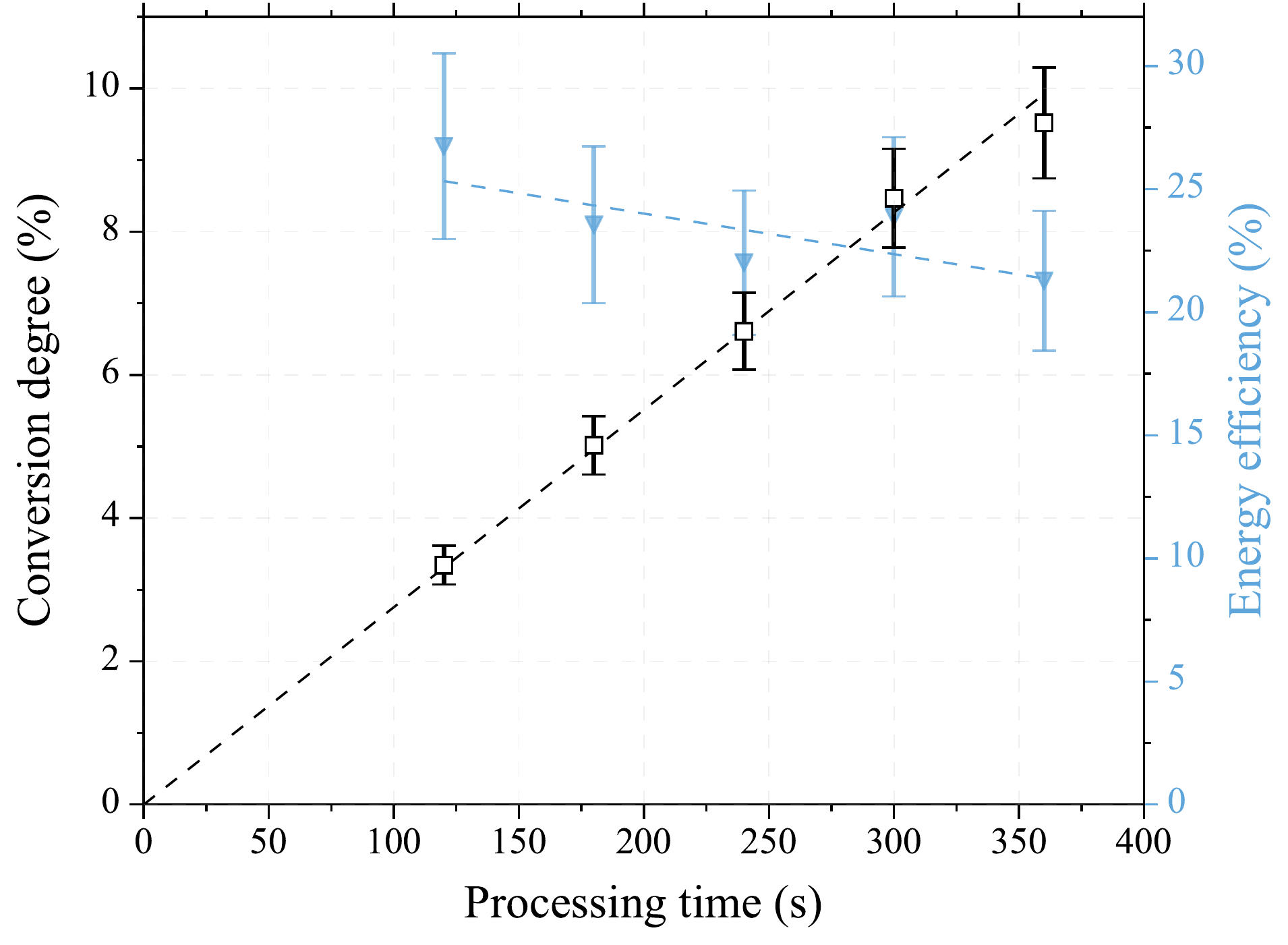}
\captionsetup{justification=centering,margin=2cm}
\caption{Influence of the processing time on the conversion degree (black) and energy efficiency (blue). Experimental conditions: p = 5 bar, f = 30 kHz, U = 800 $\mu$J.}
\label{figure5}
\end{figure}

\begin{figure}
\centering
\includegraphics[width=4in]{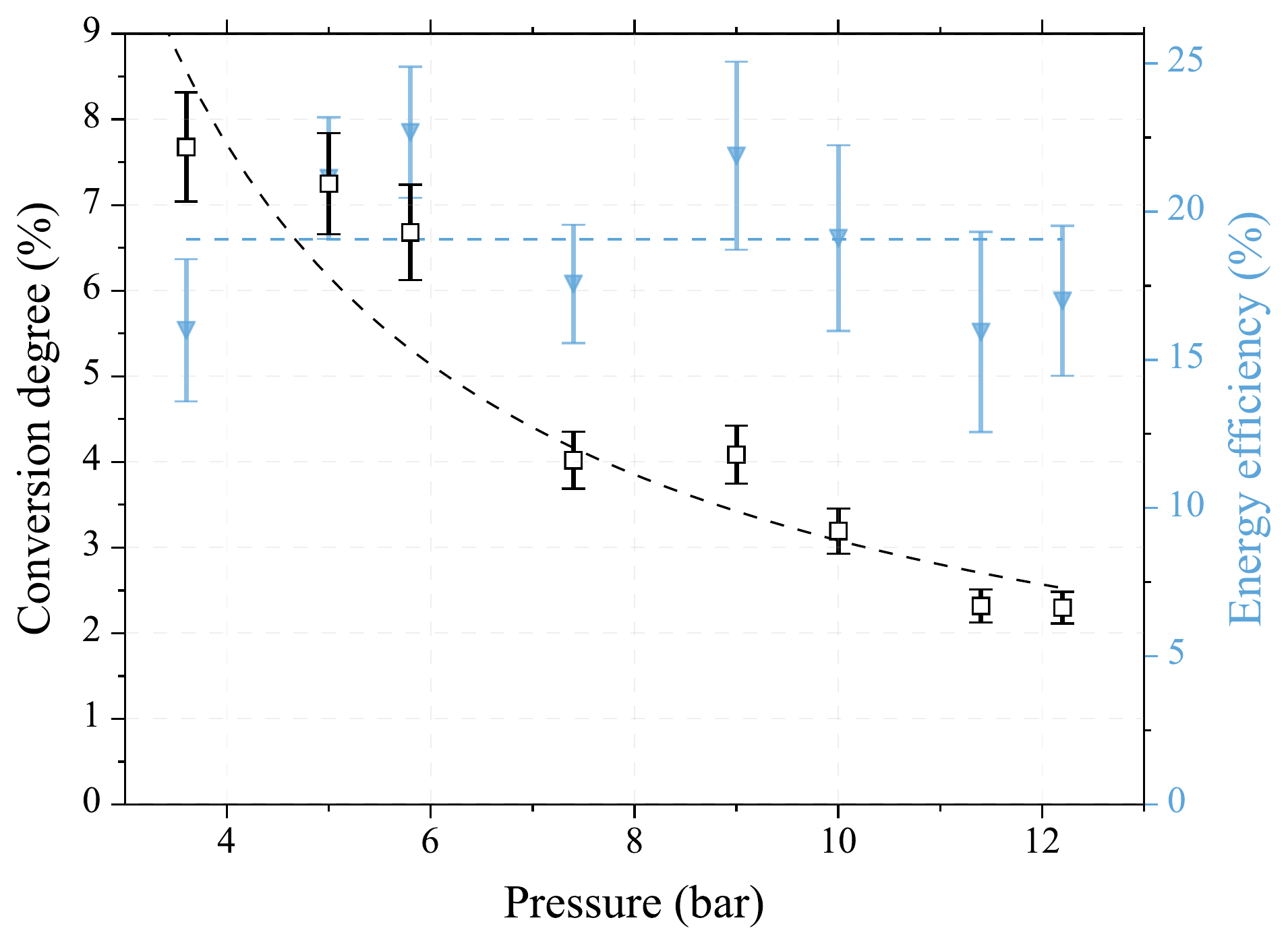}
\captionsetup{justification=centering,margin=2cm}
\caption{Influence of the pressure on the conversion degree (black) and energy efficiency (blue). E = 640 $\mu$J, $\tau$ = 400 s}
\label{figure6}
\end{figure}

\begin{figure}
\centering
\includegraphics[width=4in]{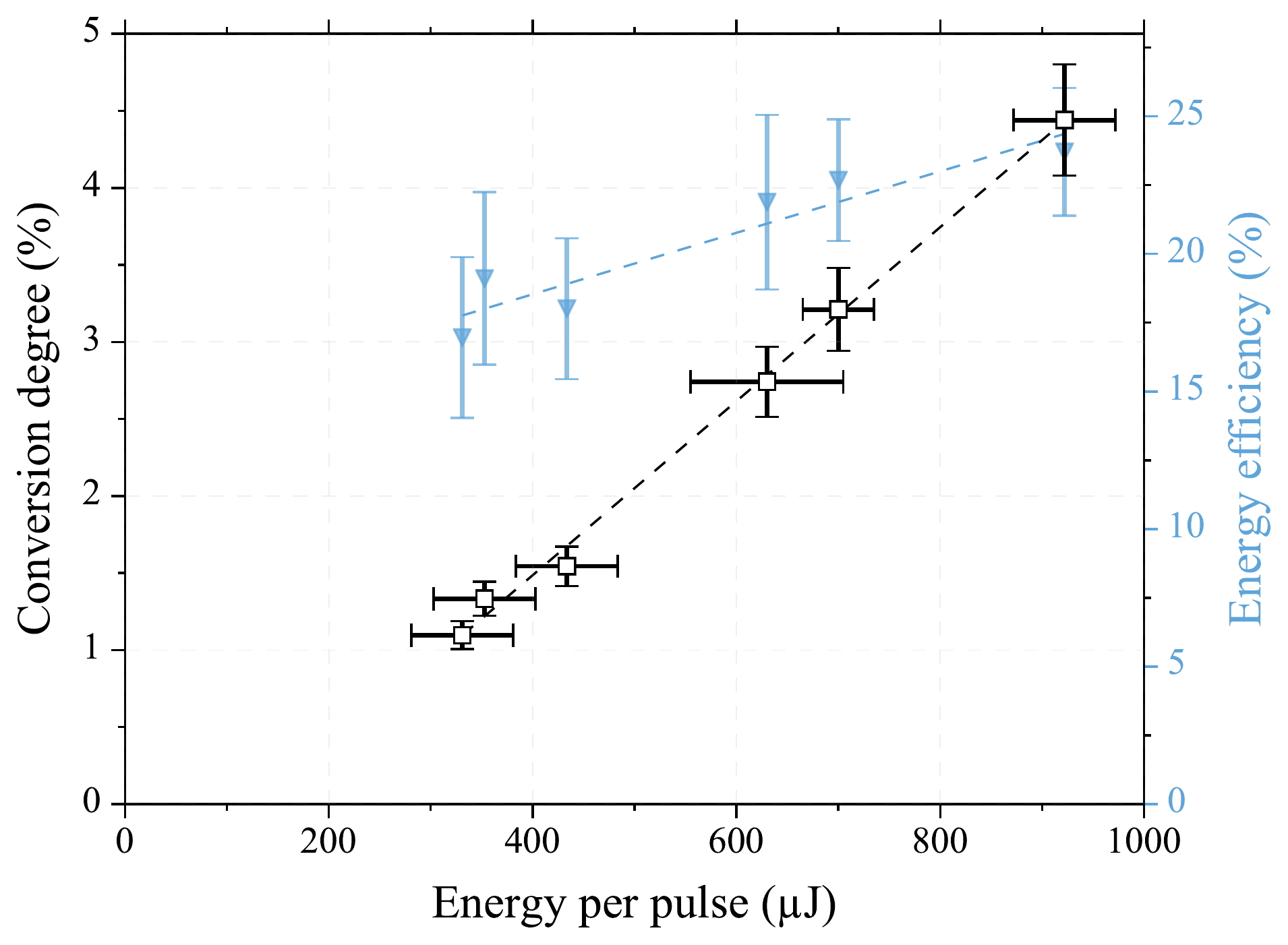}
\captionsetup{justification=centering,margin=2cm}
\caption{Influence of the energy per pulse on the conversion degree (black) and energy efficiency (blue). P=9 bar, $\tau$ =150 s.}
\label{figure7}
\end{figure}

In order to maximize the energy efficiency, the rest of the experiments are carried out while in the initial non-saturated conditions of processing time (the region in Fig.~\ref{figure4} where the products increase nearly linearly with time). In Figs.~\ref{figure5},~\ref{figure6},~\ref{figure7}, we explore the dependency of both the conversion degree and energy efficiency on the processing time, pulse energy, and chamber pressure. 
 
As expected, the conversion degree increases linearly with the processing time. The energy efficiency decreases slightly from 27 to 22\% (Fig.~\ref{figure5}). For a fixed average power, the conversion degree decreases inversely with pressure, and the energy efficiency is found to be insensitive to pressure (Fig.~\ref{figure6}). Finally, the conversion degree increases linearly with pulse energy whereas the energy efficiency increases only slightly, from 16 to 24\% (Fig.~\ref{figure7}).

Additional experiments are performed for a broad range of pressures (4-13~bar), pulse energies (300-700~$\mu$J), and processing times (40-420~s).  In all cases, the energy efficiency is close to $\eta\sim$~20\%. The slight departure from this value comes from the non-linear dependency on the pulse energy as seen in Fig.~\ref{figure7}. These results confirm that the SEI is the critical parameter that determines the conversion degree of the discharge. The highest energy efficiency is $\eta$=27\% and is obtained with a conversion rate of $\Phi$=4\% for the following conditions: P~=~5~bar, E$_{pulse}$~=~865$\pm$100~$\mu$J, $\tau$=420~s. The highest conversion rate is $\Phi$=14\% and is obtained with an energy efficiency of $\eta$=23\% for the following conditions: P~=~5.8~bar, E$_{pulse}$~=~700$\pm$75~$\mu$J, $\tau$=420~s.  

\subsection{Results in context of prior work}

\begin{figure}
\centering
\includegraphics[width=4in]{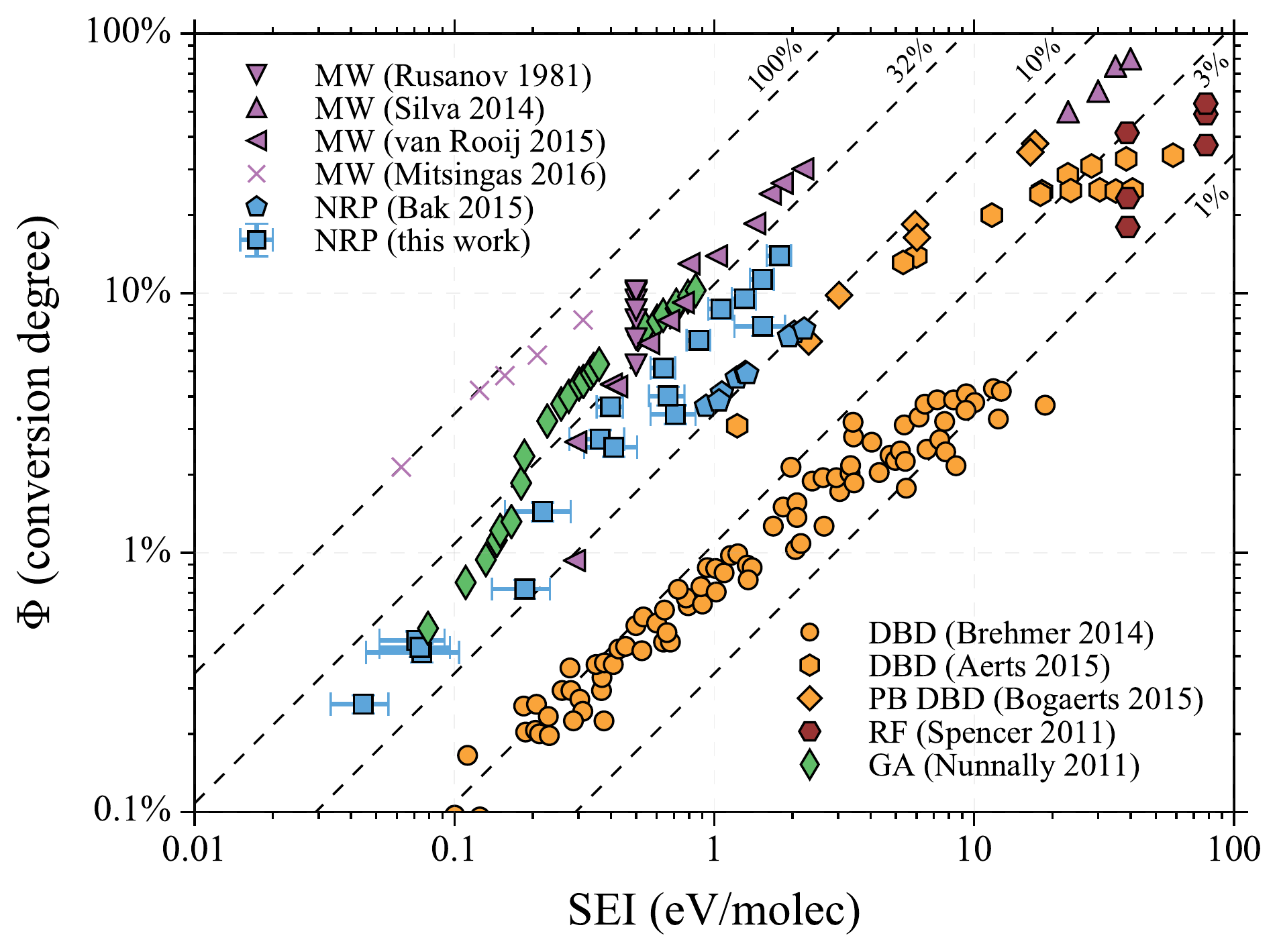}
\captionsetup{justification=centering,margin=2cm}
\caption{Conversion degree against specific energy input for different non thermal plasma techniques: microwaves (MW)~\cite{Rusanov1981,Silva2014,VanRooij2015,Mitsingas2016}, nanosecond repetitively pulsed (NRP) discharges~\cite{Bak2015a}, dielectric barrier discharges (DBD)~\cite{Brehmer2014,Aerts2015}, packed-bed (PB-DBD)~\cite{Bogaerts2015}, radio-frequency (RF)~\cite{Spencer2011d}, gliding arc~\cite{Nunnally2011}. The diagonal lines are iso-efficiency lines for energy efficiency from 1 to 100\%}
\label{figure8}
\end{figure}

Fig.~\ref{figure8} gives a summary comparison of our results with other work from the literature. It depicts the relationship between the conversion degree, energy efficiency, and the SEI for our experiments, as well as for different plasma dissociation techniques, such as DBD, packed-bed DBD, Gliding Arc, Microwave, radio frequency (RF), and other NRP discharges. Isoefficiency lines are also reported. On the logarithmic graph of conversion degree versus SEI, the energy efficiency is represented by parallel lines of slope equal to unity. 
When not directly available, SEI and conversion degrees are calculated from the reported data. The experiments of Rusanov \textit{et al.}~\cite{Rusanov1981} are carried out in a microwave discharge with pressures ranging from 50 to 300~Torr at a constant SEI. The conversion degree is derived by us from the reported energy efficiency. The work of Rooij \textit{et al.}~\cite{VanRooij2015} is also carried out using microwave plasmas, at 150~mbar, with flow rates of 5 and 15~slpm, and with varying input power. In that work, other pressures were explored, but the energy efficiencies were found to be greatest at 150~mbar. The DBD studies of Brehmer \textit{et al.}~\cite{Brehmer2014} were conducted over a range of frequencies (60-130 kHz). The microwave experiments of Silva \textit{et al.}~\cite{Silva2014} were conducted in $CO_2$ with 5\% $N_2$ addition to carry out actinometry. 

Most of the experiments reported on this graph exhibit a relatively low sensitivity of the energy efficiency on the SEI, and in particular on pressure. This finding is important to be able to scale the processes to higher throughputs and increased conversion degree without reducing the efficiency. It seems to apply particularly to some techniques. In DBD discharges, the linear dependence of the conversion degree on the SEI has already been reported by Brehmer~\textit{et al.}~\cite{Brehmer2014} for a range of density and confirmed by the analysis of Bogaerts~\textit{et al.}~\cite{Bogaerts2015}. On the other hand, low pressure discharges such as the reported microwave experiments are particularly sensitive to presure conditions as evidenced by the results of Rusanov~\textit{et al}~\cite{Rusanov1981} and of Rooij~\textit{et al.}~\cite{VanRooij2015} (although only one pressure is reported on Fig.~\ref{figure8} for the sake of clarity).

It follows that, although convenient to predict the conversion degree, the SEI is not the only parameter required to characterize the efficiency of a given technique. Experiments carried out with microwaves discharges by Rooij~\textit{et al.} and Silva~\textit{et al.} exhibit large disparities in efficiency. Similarly, in the NRP experiments of Bak~\textit{et al.} and those of this work, the energy efficiency differs by almost a factor of two (13\% to 20\% average) despite the same range of SEI investigated. In these studies, the same high voltage generator is used, hence the same pulse shape and pulse duration. The range in pressure and energy deposited overlap as well. The experiments of Bak \textit{et al.} were performed in a flow reactor, possibly reducing the processing time of a fluid element to the plasma (note that Fig.~\ref{figure4} shows that a reduced processing time has a small but positive effect on the energy efficiency). In the next section, we suggest however that the biggest impact on the difference may be due to the different reduced electric fields that were used in these two studies.

\section{Kinetics of NRP CO2 dissociation}

\subsection{Reduced electric field in a nanosecond transient discharge}

Neglecting the enhancement due to the pin-to-plane geometry, the initial electric field before breakdown in our experiments is approximated as the ratio of the applied voltage to the interelectrode gap distance:
\begin{equation}
E_0\sim\frac{V}{d}.
\label{eq15}
\end{equation}
In Bak experiments~\cite{Bak2015a}, the pressure ranges from 2.3 to 5 bar, with a voltage of 9~kV when the breakdown occurs, and a gap distance of 0.7~mm. This yields an initial reduced electric field of 105-230~Td. In our studies, the initial gap length is 150$\pm$30 $\mu$m, and increases at up to 250$\pm$30 because of ablation as stated in section~\ref{section_energy}. For our experimental result reported in Fig.~\ref{figure2} (5~bar, V=20~kV), the initial reduced electric field is in the range 660-1100~Td, which is 4 to 6 times higher than Bak experiments. After breakdown, this high reduced electric field drops quickly as the current increases, to finally reach the level obtained by an equilibrium between the rate of ionization (characterized by the ionization coefficient, $\gamma_i$), and attachment (characterized by the attachment coefficient, $\gamma_a$), as described by the following equation~\cite{Raizer}: 
\begin{equation}
\frac{\gamma_i(E/N)}{\gamma_a(E/N)}\sim 1-10.
\label{eq16}
\end{equation}
This value is independent of the initial reduced electric field and is in the range of 90-135~Td in a pure CO$_2$ mixture~\cite{Bak2015a}. This equation is valid for high currents, which is the case here as we measure conduction currents of more than 10~A. However, it describes an equilibrium in a DC regime that may not be reached during the short duration of the conduction phase (7~ns) of the NRP discharge. As a result, the effective reduced electric field that exists during the dissociation is only known to be bounded by the initial electric field of equation~(\ref{eq15}) and that described by the ionization-attachment equilibrium given by equation~(\ref{eq16}). 

\subsection{Dissociation mechanisms}

In this section we discuss the detailed reactions that may be important in CO$_2$ dissociation kinetics and discuss those that we believe to be relevant to our NRP discharge. The literature usually draws attention to five important pathways that contribute to CO$_2$ splitting. These are summarized in Table~\ref{table1}. In general, CO$_2$ splitting in plasmas requires one or more initial inelastic electron collisional processes followed by dissociation processes. The path in the table described as (Ra) is the direct electron dissociative attachment reaction that produces CO and O$^-$. Fridman~\cite{Fridman2008j} showed that step-by-step  vibrational excitation (Rv) was the most important mechanism relevant to his experiment, with the non-adiabatic transition $^1\Sigma^+ \rightarrow ^3B^2$ of highly vibrationally excited states leading to dissociation. Other important steps involve the direct electronic excitation of CO$_2$, as shown with (Re$_7$) and (Re$_{10}$), the latter of which is followed by auto dissociation.  Finally, another important path (Ri) is the ionization of CO$_2$ followed by dissociative recombination.  Cross-sections for the two electronic excitation steps, dissociative attachment, and ionization, are available from Phelps~\cite{Phelps}.  

\begin{table}[ht!]
  \centering
  \caption{Possible kinetic mechanisms involved in CO$_2$ splitting}
  \label{tab:table1}
  \begin{tabular}{|l|p{13cm}|}
  \hline
 & Dissociative attachment (3.85~eV): \\
(Ra)	 & $e$+CO$_2$ $\rightarrow$ CO+O$^-$\\
 \hline
 & Step-by-step vibrational excitation:\\
(Rv)	 & $e$+CO$_2$ $\rightarrow$ $e$+CO$_2$(v*) \;\;\;\;\;\;\;\;\;\;\;\;\;\;\;\; CO$_2$(v*) $\rightarrow$ CO$_2$($^3B_2$) $\rightarrow$  CO+O \\
 \hline
 & Direct electronic excitation followed by auto-dissociation:\\ 
(Re$_7$)	  & $e$+CO$_2$ $\rightarrow$ $e$+CO$_2$* (7~eV) \,\;\;\;\;\;\;\;\;\,\, CO$_2$* $\rightarrow$ CO+O \\
(Re$_{10}$)	   & $e$+CO$_2$ $\rightarrow$ $e$+CO$_2$* (10.5~eV) \;\;\;\;\;\,\, CO$_2$* $\rightarrow$ CO+O \\
 \hline
 & Ionization (13.3~eV) followed by dissociative recombination: \\
(Ri)	 & $e$+CO$_2\rightarrow 2e+CO_2^+$\;\;\;\;\;\;\;\;\;\;\;\;\;\;\;\;\;\;\;\; $e$+CO$_2^+$ $\rightarrow$ CO+O\\
 \hline
  \end{tabular}
\label{table1}
\end{table}

\begin{figure}
\centering
\includegraphics[width=4in]{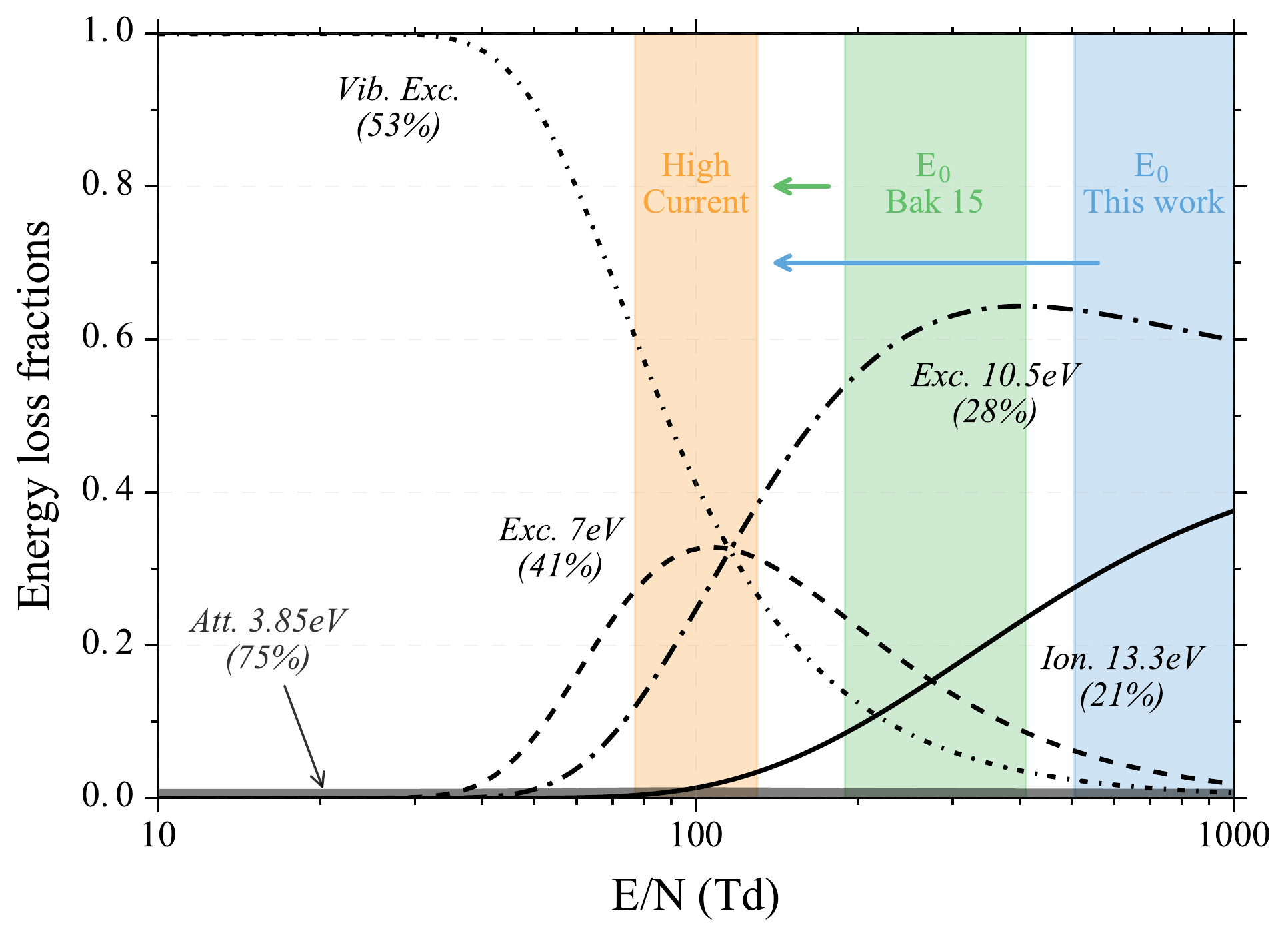}
\captionsetup{justification=centering,margin=2cm}
\caption{Energy losses fractions for different electron collisions (attachement, sum of vibrational excitation, electronic excitation, ionization), and upper and lower bounds of the reduced electric field (E/N) in~\cite{Bak2015a} and this work. The ratio of the dissociation enthalpy over the reaction energy is shown for each electron collision.}
\label{figure9}
\end{figure}

Fig.~\ref{figure9} gives the fraction of the electron energy lost to the different CO$_2$ excitation and ionization processes described in the table. These calculations are facilitated by the use of Bolsig+~\cite{Hagelaar2005c}, a software tool that solves the electron Boltzmann equation for a given reduced electric field with a specified database for electron-CO$_2$ collisional cross-sections~\cite{Phelps}. In the figure, we depict the estimated range (lower and upper bound) of the reduced electric field of our experiments (discussed in the previous section) as well as that of the experiments of Bak~\cite{Bak2015a}, which, as mentioned above, were carried out under a similar discharge configuration but at lower pressure.

The electron energy lost in the inelastic collisions described in Table~\ref{table1} can be used, together with the enthalpy of the CO$_2$ dissociation reaction (2.9~eV), to determine a possible upper bound for the energy efficiency of each mechanism. The result is considered an upper bound because we neglect other reactions, if any, that compete with the subsequent step, such as auto-dissociation or dissociative recombination.  The dissociative attachment process (Ra) is a fairly efficient process but very marginal at high reduced electric fields. Because direct electronic excitation (7~eV, 10.5~eV) and ionization (13.3~eV) have an energy threshold much higher than the CO=O bond energy (5.5~eV), the resulting energy efficiency is bounded by 41\%, 28\% and 21\%, respectively. Finally, the energy needed for dissociation by step-by-step vibrational dissociation is exactly the CO=O bond energy, which yields a bound on the efficiency of 53\%. We note however, that caution should be used in interpreting these bounds as absolute. The energy converion efficiencies can be increased further whenever an oxygen atom that is produced by dissociation, is consumed by the recycling reaction suggested in the macroscopic model:
$$
CO_2 + O \rightarrow CO + O_2
$$
This reaction, which  is almost iso-energetic, can lead to a significant increase in energy efficiency, if it occurs at a rate faster than direct oxygen recombination (which is strongly exothermic and leads to local gas heating). This may explain why energy efficiencies higher than 53\% have already been reported in vibrationally hot plasmas~\cite{Rusanov1981}. This enhancement is expected to be independent of the mechanism that leads to the production of atomic oxygen; \textit{i.e}, vibrational and electronic mechanisms will both benefit from it. Below, we shall assume that this recycling mechanism is unimportant (the macroscopic model introduced earlier to describe the experiment suggest that its rate is at least a factor of two lower than the recombination reaction of O with CO), and furthermore discuss how inelastic electron collisions are believed to play the most significant role in NRP dissociation kinetics.

\subsection{Mechanisms in NRP discharges}

The high electric field in NRP experiments ($>$100 Td) inhibit the development of the so-called vibrational-ladder mechanism. If this vibrational excitation mechanism was the only reaction at play in dissociation, decreasing the reduced electric field ould yield higher efficiencies as more of the electron energy would be used to excite vibrational modes. However, we observed the opposite trend. The experiments of Bak~\textit{et al}, that were performed at a lower reduced electric field yielded a lower efficiency. The parametric study of the influence of variations in pulse energy on efficiency (Fig.~\ref{figure7}) shows similar results as the energy efficiency increases slightly with increases in pulse energy. 

\begin{figure}
\centering
\includegraphics[width=6in]{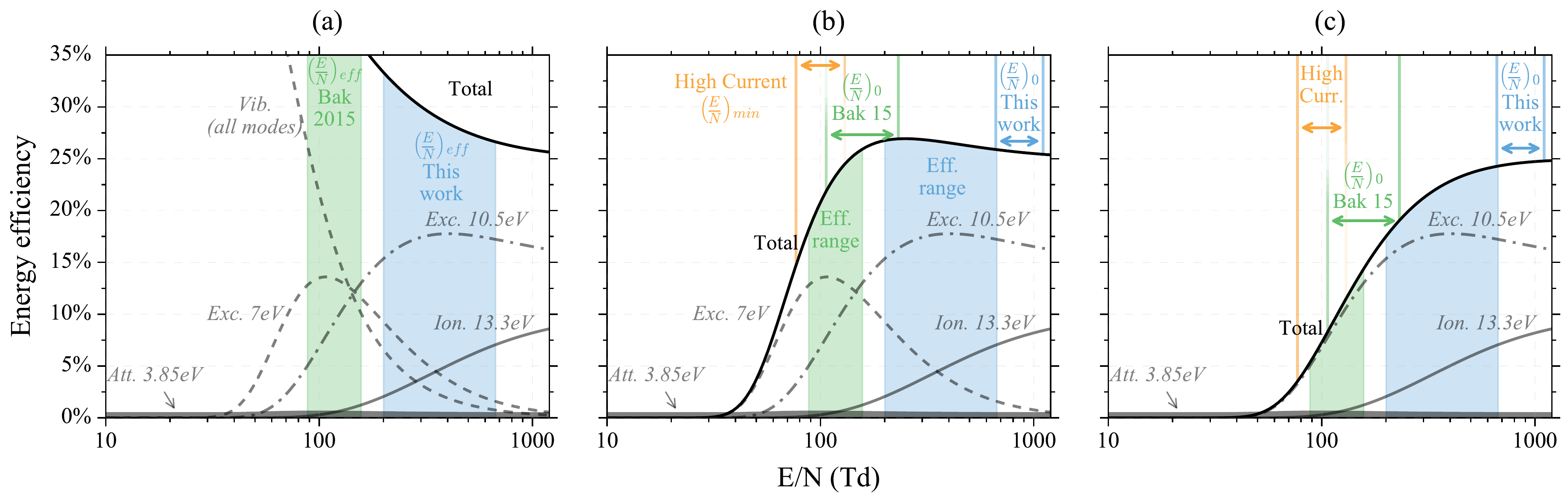}
\captionsetup{justification=centering,margin=2cm}
\caption{Contribution to the energy efficiency of the different mechanisms. (a) All mechanisms. (b) No vibrational excitation. (c) No vibrational excitation, no 7~eV electronic excitation. Effective field is bounded by initial field and high current equilibrium value. The recycling of atomic oxygen may increase the overall efficiency but is assumed not to change the relative contribution of the different mechanisms (more details in text)}
\label{figure10}
\end{figure}

Fig.~\ref{figure10} describes the contribution of the various reactions presented in Table I to the upper bound for the energy efficiency. The contribution of each dissociation path is determined by multiplying the energy loss fraction (previously shown in Fig.~\ref{figure9}) with the ratio of the dissociation enthalpy to the energy required for the particular reaction. We assume that the oxygen atom recycling reaction is negligible. All mechanisms are considered in Fig.~\ref{figure10} (a). In (b), the vibrational mechanism (Rv) is neglected. In (c), the (Re$_7$) electronic excitation is also neglected. Fig.~\ref{figure10} also shows the lower and upper bounds of the reduced electric field for Bak and our experiments, as well as the effective reduced electric field as a qualitative range between these two extrema. 

For case (a), it appears that the  energy efficiency (described by the solid line) would decrease with increases in the reduced electric field. For case (b), it would be almost constant for E/N$>$100 Td, which is the expected range in the experiments of Bak and our current studies. Finally, for case (c) there is a clear increase in the energy efficiency with increases in the reduced electric field. Since the measured energy efficiencies of about 20\% are substantially greater than those of Bak (13\%), we believe that case (c) reflects the conditions at play in our studies, and hypothesize the following:

\begin{itemize}

\item The vibrational mechanism may not be important in nanosecond pulsed discharges, even at lower values of the reduced electric field. We believe that the short pulse duration prevents the vibrational-ladder mechanism that may be at play in other discharges. Although the vibrational excitation rate in a 10~bar CO$_2$ mixture (estimated with Bolsig+~\cite{Hagelaar2005c}) yields a characteristic time for excitation of the lowest vibrational levels (\~10$^{-12}$~s) that is short in comparison to the pulse duration, excitation to the upper vibrational levels (where dissociation occurs), requires vibrational-vibrational (V-V) transfer between excited CO$_2$ molecules that is much slower~\cite{Fridman2008j}, occurring over times that may be longer than the pulse duration. 

\item We also believe that the (Re$_7$) electronic excitation process may not contribute strongly to the dissociation process. The energy involved in this reaction (7~eV) is higher than the CO=O bond enthalpy (532~kJ/mol, 5.5~eV), but the activation energy for this dissociation pathway may be higher. If this low contribution is a specificity of NRP discharges, it may also be explained in the context of the multistep dissociation mechanism whose initial or later stages is limited by the short pulse duration.  

\end{itemize}

\section{Conclusion}

In this work we have investigated the dissociation of high pressure CO$_2$ with nanosecond pulse discharges in a batch reactor configuration. For conditions where the conversion degree is less than 15\%, the energy efficiency is found to be about 20\% and exhibits a low sensitivity to the pressure, pulse energy and residence time conditions. The energy conversion degree can be expressed as a linear function of the specific energy input (SEI). A conversion rate of up to $\Phi$=14\% is achieved with an energy efficiency of $\eta$=23\% for the following conditions: P~=~5.8~bar, E$_{pulse}$~=~700$\pm$75~$\mu$J, $\tau$=420~s.  

Comparing our results with previous NRP experiments, we find the energy efficiency to be almost twice that of these previous studies. We attributed this finding to the higher electric fields that are used in this work. An examination of  the different CO$_2$ dissociation mechanisms suggests that the vibrational mechanism often responsible for the levels of dissociation seen in other plasma devices is not likely to be significant in these nanosecond pulses,  as our results show that a lowering of the reduced electric field reduces the measured energy efficiency.

\section*{Appendix A: Electro-chemical etching of the electrodes }
\setcounter{equation}{0}
\renewcommand{\theequation}{A.\arabic{equation}}

In this Appendix we describe the chemical process used to reproducibly fabricate our electrodes. It is a variant of the process first developed by Hobara \textit{et al}~\cite{Hobara2007} for the fabrication of tips used in multi-tip scanning tunnelling microscopes. A 2~mm wide tungsten wire is first fastened onto a holder that is connected to a vertical translation pod and then submerged into a solution of sodium hydroxide NaOH (2~mol/L). This submerged wire will serve as an electrochemical anode. The portion of the wire that is not submerged is connected to the positive potential side of a DC (voltage regulated) power supply. The grounded lead of the power supply is connected to a second electrode that is also submerged into the solution. This second electrode serves as a cathode. The length (L) of the submerged portion influences the radius of curvature of the etched electrode, which is a consequence of the electrochemistry that occurs at the region of the submerged wire near the free surface (described below). We used L=6~mm to obtain a radius of curvature of 10~$\mu$m when 20 V is applied to the wire. 
The length of the electrode following the etch process, as subsequently used in the discharge is the non submerged part of the tungsten wire, and is measured with the vertical translation stage. When installed in the reactor, the electrode gap distance is measured with a precision of approximately 30 $\mu$m.

\begin{figure}
\centering
\includegraphics[width=4in]{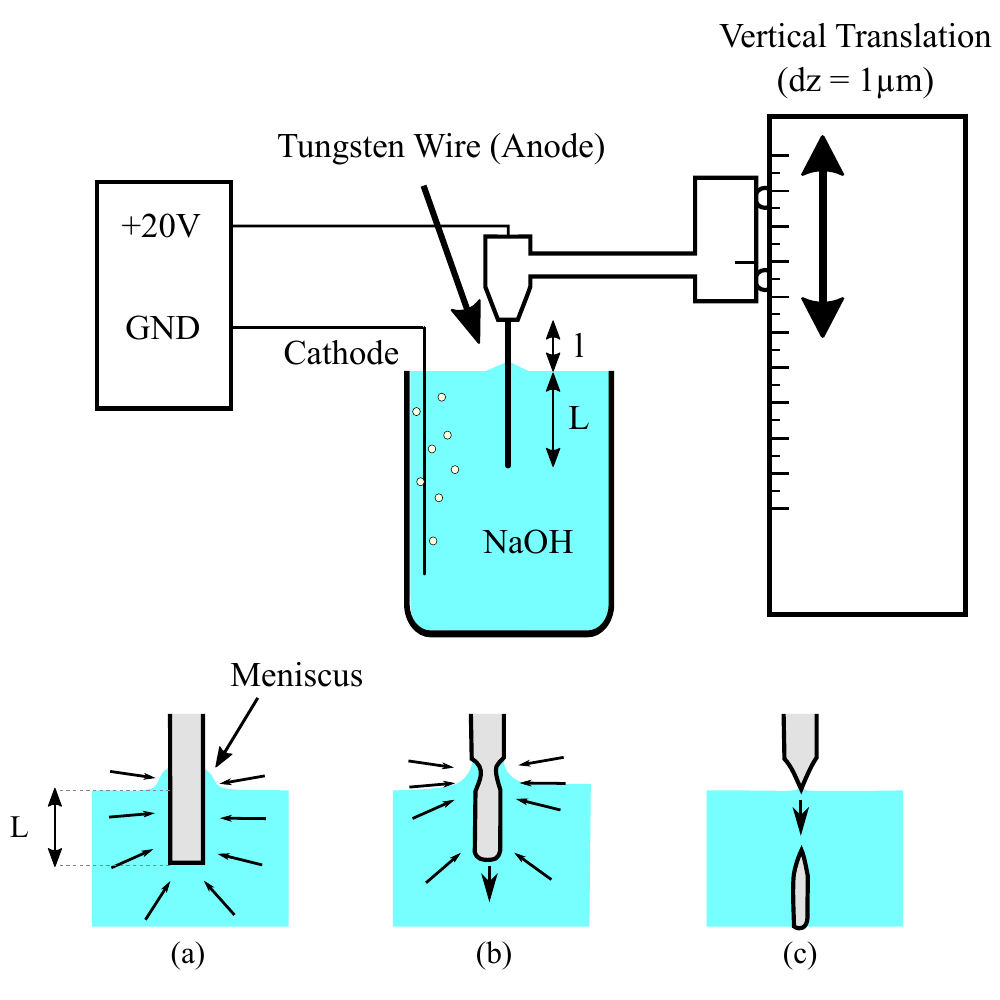}
\captionsetup{justification=centering,margin=2cm}
\caption{Electrodes etching process}
\label{figure13}
\end{figure}

\begin{figure}
\centering
\includegraphics[width=2in]{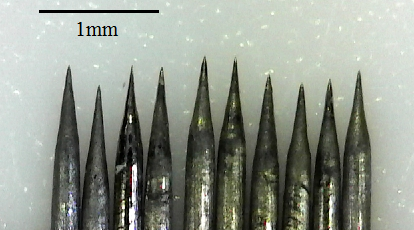}
\captionsetup{justification=centering,margin=2cm}
\caption{Tungsten electrodes manufactured with NaOH etching. Radius of curvature R\textless 10~$\mu$m}
\label{figure14}
\end{figure}

The resulting anodic reactions are described by a six-step mechanism~\cite{Kelsey1977}. The global reaction is represented as $W$+8OH$^-$ $\rightarrow$ 4H$_2$O+WO$_4^{-}$+6$e$, leading to the formation of aqueous tungsten oxide ions and the liberation of tungsten from the surface.
The rate of electrochemical etching is faster just below the meniscus, as the convection of OH$^-$ ions increases the oxidation rate of tungsten~\cite{Hobara2007} (Fig.~\ref{figure13}). As the wire diameter below the meniscus becomes extremely small, the weight of the residual immersed part stretches the tip of the future electrode. Eventually, the lower part is detached under the influence of the combination of anodic oxidation and gravity resulting in a sharp electrode tip. The entire etching process is complete in approximately 20~s.  The process results in a consistent taper and tip radius of curvature, as shown by the comparison of ten fabricated samples in the photograph of Fig.~\ref{figure14}.

\section*{Appendix B: Analysis of gas chromatographic samples}
\setcounter{equation}{0}
\renewcommand{\theequation}{B.\arabic{equation}}

In this appendix we discuss the operation of the gas chromatograph and its use for measuring CO and O$_2$ mole fractions. Two main hypotheses are made in the interpretation of the signals:

\begin{enumerate}[(i)]
\item We assume the mixture is inert after the end of the experiment. Successive measurements were carried out with 10~min interval times and no statistically significant change in the chemical composition could be observed, validating this assumption. Chemical reactions inside the GC are also neglected. In particular, we neglect surface CO recombination that may happen in the narrow GC tubing.  
\item We assume CO and CO$_2$ are the only two carbon compounds created. The carbon balance equation is then $N_{CO}+N_{CO_2}=N^o$. This assumption is confirmed by our measurements, as  no other C$_1$ or any C$_2$ species are detected, despite the sensitivity of the PDHID detector to all carbon-based compounds. 
\end{enumerate}

In extracting samples from the reactor, a mass flow controller (MFC) is used between the chamber and the GC to maintain a constant flow of 50~cc/min to the detector. 

The calibrated GC signal is proportional to the mole fraction of each species. Under (i), the mole fractions in the chamber and in the GC columns are preserved. The molar ratio in the chamber are derived directly from the GC signals. Carbon atom balance results in the following expression for the conversion degree :
\begin{equation}
\Phi = \frac{N_{CO}}{N^0}=\frac{\chi_{CO}}{\chi_{CO_2}+\chi_{CO}}
\label{19}
\end{equation}
$N^0$ is determined from the initial chamber pressure, $P_0$ , the chamber volume, $V$, and the CO$_2$ compressibility, $Z(P_0)$. The number of CO molecules produced in the reactor is then expressed as:
\begin{equation}
N_{CO} = \Phi\cdot N^0=\frac{\chi_{CO}}{\chi_{CO_2}+\chi_{CO}}\cdot\frac{P_0V}{Z({P_0})RT}
\label{20}
\end{equation}

It is noteworthy that neglecting CO$_2$ compressibility would significantly underestimate the number of CO molecules created in the chamber.  For chamber pressures P $>$10 bar, this error would be greater than 5\%.

\section*{Acknowledgment}
 
Taemin Wong acknowledges support from the Precourt Institute for Energy and the TomKat Center for Sustainable Energy.  E.P. acknowledges the  France-Stanford Center for Interdisciplinary Studies, which provided the funding for his Stanford Visiting Student appointment. 

\section*{References}

\bibliography{biblio}

\begin{thebibliography}{10}

\bibitem{Lebouvier2013}
Lebouvier A, Iwarere SA, d’Argenlieu P, Ramjugernath D, Fulcheri L.
\newblock Assessment of carbon dioxide dissociation as a new route for syngas
  production: a comparative review and potential of plasma-based technologies.
\newblock Energy \& Fuels. 2013;27(5):2712-22.

\bibitem{Goeppert2014}
Goeppert A, Czaun M, Jones JP, Prakash GS, Olah GA.
\newblock Recycling of carbon dioxide to methanol and derived products--closing
  the loop.
\newblock Chemical Society Reviews. 2014;43(23):7995-8048.

\bibitem{Petitpas2007}
Petitpas G, Rollier JD, Darmon A, Gonzalez-Aguilar J, Metkemeijer R, Fulcheri
  L.
\newblock A comparative study of non-thermal plasma assisted reforming
  technologies.
\newblock International Journal of Hydrogen Energy. 2007;32(14):2848-67.

\bibitem{Rusanov1981}
Rusanov VD, Fridman A, Sholin GV.
\newblock The physics of a chemically active plasma with nonequilibrium
  vibrational excitation of molecules.
\newblock Soviet Physics Uspekhi. 1981;24(6):447.

\bibitem{VanRooij2015}
Van~Rooij G, Van Den~Bekerom D, Den~Harder N, Minea T, Berden G, Bongers W,
  et~al.
\newblock Taming microwave plasma to beat thermodynamics in CO 2 dissociation.
\newblock Faraday discussions. 2015;183:233-48.

\bibitem{Goede2014}
Goede AP, Bongers WA, Graswinckel MF, van~de Sanden RM, Leins M, Kopecki J,
  et~al.
\newblock Production of solar fuels by CO2 plasmolysis.
\newblock In: EPJ web of conferences. vol.~79. EDP Sciences; 2014. p. 01005.

\bibitem{Kozak2015}
Koz{\'a}k T, Bogaerts A.
\newblock Evaluation of the energy efficiency of CO2 conversion in microwave
  discharges using a reaction kinetics model.
\newblock Plasma Sources Science and Technology. 2014;24(1):015024.

\bibitem{Spencer2011d}
Spencer LF, Gallimore AD.
\newblock Efficiency of CO2 dissociation in a radio-frequency discharge.
\newblock Plasma Chemistry and Plasma Processing. 2011;31(1):79-89.

\bibitem{Wu1996}
Wu D, Outlaw R, Ash R.
\newblock Extraction of oxygen from CO2 using glow-discharge and permeation
  techniques.
\newblock Journal of Vacuum Science \& Technology A: Vacuum, Surfaces, and
  Films. 1996;14(2):408-14.

\bibitem{Wang1999}
Wang JY, Xia GG, Huang A, Suib SL, Hayashi Y, Matsumoto H.
\newblock CO2 decomposition using glow discharge plasmas.
\newblock Journal of catalysis. 1999;185(1):152-9.

\bibitem{Spencer2012a}
Spencer L, Gallimore A.
\newblock CO2 dissociation in an atmospheric pressure plasma/catalyst system: a
  study of efficiency.
\newblock Plasma Sources Science and Technology. 2012;22(1):015019.

\bibitem{Mitsingas2016}
Mitsingas CM, Rajasegar R, Hammack S, Do H, Lee T.
\newblock High energy efficiency plasma conversion of CO 2 at atmospheric
  pressure using a direct-coupled microwave plasma system.
\newblock IEEE Transactions on Plasma Science. 2016;44(4):651-6.

\bibitem{Paulussen2010}
Paulussen S, Verheyde B, Tu X, De~Bie C, Martens T, Petrovic D, et~al.
\newblock Conversion of carbon dioxide to value-added chemicals in atmospheric
  pressure dielectric barrier discharges.
\newblock Plasma Sources Science and Technology. 2010;19(3):034015.

\bibitem{Brehmer2014}
Brehmer F, Welzel S, Van De~Sanden M, Engeln R.
\newblock CO and byproduct formation during CO2 reduction in dielectric barrier
  discharges.
\newblock Journal of Applied Physics. 2014;116(12):123303.

\bibitem{Aerts2015}
Aerts R, Somers W, Bogaerts A.
\newblock Carbon dioxide splitting in a dielectric barrier discharge plasma: a
  combined experimental and computational study.
\newblock ChemSusChem. 2015;8(4):702-16.

\bibitem{VanLaer2015}
Van~Laer K, Bogaerts A.
\newblock Improving the conversion and energy efficiency of carbon dioxide
  splitting in a zirconia-packed dielectric barrier discharge reactor.
\newblock Energy Technology. 2015;3(10):1038-44.

\bibitem{Mei2015d}
Mei D, Zhu X, He YL, Yan JD, Tu X.
\newblock Plasma-assisted conversion of CO2 in a dielectric barrier discharge
  reactor: understanding the effect of packing materials.
\newblock Plasma Sources Science and Technology. 2014;24(1):015011.

\bibitem{Zeng2015c}
Zeng Y, Zhu X, Mei D, Ashford B, Tu X.
\newblock Plasma-catalytic dry reforming of methane over $\gamma$-Al2O3
  supported metal catalysts.
\newblock Catalysis Today. 2015;256:80-7.

\bibitem{Snoeckx2016}
Snoeckx R, Heijkers S, Van~Wesenbeeck K, Lenaerts S, Bogaerts A.
\newblock CO 2 conversion in a dielectric barrier discharge plasma: N 2 in the
  mix as a helping hand or problematic impurity?
\newblock Energy \& Environmental Science. 2016;9(3):999-1011.

\bibitem{Ramakers2015d}
Ramakers M, Michielsen I, Aerts R, Meynen V, Bogaerts A.
\newblock Effect of argon or helium on the CO2 conversion in a dielectric
  barrier discharge.
\newblock Plasma Processes and Polymers. 2015;12(8):755-63.

\bibitem{Indarto2007}
Indarto A, Yang DR, Choi JW, Lee H, Song HK.
\newblock Gliding arc plasma processing of CO2 conversion.
\newblock Journal of hazardous materials. 2007;146(1-2):309-15.

\bibitem{Nunnally2011}
Nunnally T, Gutsol K, Rabinovich A, Fridman A, Gutsol A, Kemoun A.
\newblock Dissociation of CO2 in a low current gliding arc plasmatron.
\newblock Journal of Physics D: Applied Physics. 2011;44(27):274009.

\bibitem{Tu2014}
Tu X, Whitehead JC.
\newblock Plasma dry reforming of methane in an atmospheric pressure AC gliding
  arc discharge: Co-generation of syngas and carbon nanomaterials.
\newblock International journal of hydrogen energy. 2014;39(18):9658-69.

\bibitem{Scapinello2016}
Scapinello M, Martini L, Dilecce G, Tosi P.
\newblock Conversion of CH4/CO2 by a nanosecond repetitively pulsed discharge.
\newblock Journal of Physics D: Applied Physics. 2016;49(7):075602.

\bibitem{Bak2015a}
Bak MS, Im SK, Cappelli M.
\newblock Nanosecond-pulsed discharge plasma splitting of carbon dioxide.
\newblock IEEE Transactions on Plasma Science. 2015;43(4):1002-7.

\bibitem{Fridman2008j}
Fridman A.
\newblock Plasma chemistry.
\newblock Cambridge university press; 2008.

\bibitem{Rusterholtz2013}
Rusterholtz D, Lacoste D, Stancu G, Pai D, Laux C.
\newblock Ultrafast heating and oxygen dissociation in atmospheric pressure air
  by nanosecond repetitively pulsed discharges.
\newblock Journal of Physics D: Applied Physics. 2013;46(46):464010.

\bibitem{Pai2010r}
Pai DZ, Lacoste DA, Laux CO.
\newblock Nanosecond repetitively pulsed discharges in air at atmospheric
  pressure—the spark regime.
\newblock Plasma Sources Science and Technology. 2010;19(6):065015.

\bibitem{Silva2014}
Silva T, Britun N, Godfroid T, Snyders R.
\newblock Optical characterization of a microwave pulsed discharge used for
  dissociation of CO2.
\newblock Plasma Sources Science and Technology. 2014;23(2):025009.

\bibitem{Bogaerts2015}
Bogaerts A, Koz{\'a}k T, Van~Laer K, Snoeckx R.
\newblock Plasma-based conversion of CO 2: current status and future
  challenges.
\newblock Faraday discussions. 2015;183:217-32.

\bibitem{Raizer}
Raizer YP, Allen JE.
\newblock Gas discharge physics. vol.~1.
\newblock Springer; 1991.

\bibitem{Phelps}
Hake RD, Phelps AV.
\newblock Momentum-Transfer and Inelastic-Collision Cross Sections for
  Electrons in ${\mathrm{O}}_{2}$, CO, and C${\mathrm{O}}_{2}$.
\newblock Phys Rev. 1967 Jun;158:70-84.
\newblock Available from:
  \url{https://link.aps.org/doi/10.1103/PhysRev.158.70}.

\bibitem{Hagelaar2005c}
Hagelaar G, Pitchford LC.
\newblock Solving the Boltzmann equation to obtain electron transport
  coefficients and rate coefficients for fluid models.
\newblock Plasma sources science and technology. 2005;14(4):722.

\bibitem{Hobara2007}
Hobara R, Yoshimoto S, Hasegawa S, Sakamoto K.
\newblock Dynamic electrochemical-etching technique for tungsten tips suitable
  for multi-tip scanning tunneling microscopes.
\newblock e-Journal of Surface Science and Nanotechnology. 2007;5:94-8.

\bibitem{Kelsey1977}
Kelsey GS.
\newblock The anodic oxidation of tungsten in aqueous base.
\newblock Journal of the Electrochemical Society. 1977;124(6):814.

\end{thebibliography}

\end{document}